# Study of molecular spin-crossover complex $Fe(phen)_2(NCS)_2$ thin films

S. Shi [a], G. Schmerber, J. Arabski, J.-B. Beaufrand, D. J. Kim, S. Boukari, M. Bowen, N. T. Kemp, N. Viart, G. Rogez, and E. Beaurepaire [b]

*IPCMS UMR 7504 CNRS - Université de Strasbourg, 23 rue du Loess, BP 43, 67034 Strasbourg Cedex 2, France*

H. Aubriet, J. Petersen, C. Becker, D. Ruch

*LTI, Centre de Recherche Public Henri Tudor, 66 rue du Luxembourg, Esch/Alzette 4002, Luxembourg*

We report on the growth by evaporation under high vacuum of high-quality thin films of $Fe(phen)_2(NCS)_2$ (phen=1,10-phenanthroline) that maintain the expected electronic structure down to a thickness of 10 nm and that exhibit a temperature-driven spin transition. We have investigated the current-voltage characteristics of a device based on such films. From the space charge-limited current regime, we deduce a mobility of $6.5\times10^{-6}$ $cm^2/V\cdot s$ that is similar to the low-range mobility measured on the widely studied tris(8-hydroxyquinoline)aluminium organic semiconductor. This work paves the way for multifunctional molecular devices based on spin-crossover complexes.

a) shi@ipcms.u-strasbg.fr
b) eric.beaurepaire@ipcms.u-strasbg.fr

The molecular spin state of spin-crossover (SCO) complexes can be switched between a high-spin (HS) and a low-spin (LS) state by various perturbations, including temperature, optical excitation and pressure. This class of organic semiconductor materials may thus play a potentially significant role in the emerging fields of molecular electronics and nanotechnologies.[1] The thermal hysteresis in the spin transition of several types of SCO complexes[1, 2] is in particular expected to yield important applications in organic memory devices.

However, most work on SCO complexes is performed on powder samples, and as such is incompatible with industrial needs for thin films with a precise thickness control. Until now, only a few methods have successfully yielded thin films of SCO complexes: Langmuir–Blodgett (LB) deposition,[3] molecular self-assembly,[4] dip coating/drop casting,[5] spin-coating,[6] dispersing complexes into a polymer matrix and intercalating complexes in layered materials host matrixes.[7,8]

The LB method can provide very good homogeneity and thickness control of the order of one monolayer, but as a disadvantage requires the presence of special functional groups in the SCO, and is therefore materials-dependent.[3] Grid-type complexes can align along the step edges of the highly-oriented pyrolytic graphite (HOPG) surface to form a densely-packed film, but this approach is currently limited to HOPG substrates, which are expensive.[4] The dip coating/drop casting method is both simple and applicable to all soluble complexes, but it yields inhomogeneous, micron-thick films.[5] Spin-coating is a simple, widely-used method in organic/polymer electronics,[9] but there are very few reports on films obtained by spin-coating because the solubility of most SCO complexes in standard solvents is low.[6] SCO complexes have also been inserted into layered materials or plates ($CdPS_3$, $MnPS_3$, etc…) to make thin films, yet these films are unstable over time and contain remnant solvents that can affect the spin transition properties.[8]

Evaporation under high vacuum is another method used in organic/polymer electronics.[10] It produces high-quality films with a precise control of the thickness and morphology, and can be fully integrated within routine thin-film elaboration methods used in electronics. However, there is no report on SCO complexes thin films fabrication by this method.

In this letter, we report on high-quality thin films of $Fe(phen)_2(NCS)_2$ [(phen=1,10-phenanthroline); see inset of Fig. 2(a)] that were deposited by evaporation under high vacuum. The complex was synthesized according to a previous described method.[11] Thin films of $Fe(phen)_2(NCS)_2$ were deposited on silicon or glass substrates by thermal evaporation from a molybdenum boat in a chamber with a base pressure of $10^{-8}$ mbar. The deposition rate was 0.1 nm/s. The thickness was measured *in-situ* by a quartz crystal thickness monitor, and *ex-situ* by a stylus profilometer and by X-ray reflectivity.

Fig. 1 shows images obtained by atomic force microscopy (AFM) on $Fe(phen)_2(NCS)_2$ thin films fabricated by high vacuum evaporation. The rms roughness is 0.33 nm for the 280 nm-thick film on silicon and 0.29 nm for the 240 nm-thick film on glass. For both silicon and glass substrates, the films are smooth with no visible grains. However, the film morphology is sensitive to the environment and becomes coarse when exposed to air. When necessary, we have therefore encapsulated our films.

To demonstrate that the molecule remains intact after evaporation, we performed X-ray photoelectron spectroscopy (XPS) on a 280 nm-thick film of $Fe(phen)_2(NCS)_2$. XPS analyses were performed in a vacuum of $1\times10^{-9}$ mbar using a monochromatized $AlK_\alpha$ source working at 1486.7eV. A charge compensator was also used during acquisition. $Ar^+$ etching at 1keV was carried out at a pressure of $5\times10^{-5}$ mbar to eliminate a few nanometers of possible surface contaminants from the sample.

We present in Fig. 2(a) XPS spectra at the N 1*s* edge. These spectra exhibit two neighboring peaks with an intensity ratio 2:1 that reflects the nitrogen sites in the thiocyanate (at 397.6 eV) and in the 1,10-phenanthroline ligands (at 399.6 eV), [12] as expected from the $Fe(phen)_2(NCS)_2$ formula [see the inset of Fig. 2 (a)]. XPS spectra at the Fe $2p_{3/2}$ and Fe $2p_{1/2}$ edges for this film are shown in Fig. 2 (b). Between the Fe $2p_{3/2}$ and Fe $2p_{1/2}$ peaks, there is a shoulder in the spectrum (see arrow) that is characteristic of $Fe^{2+}$ in the HS state of $Fe(phen)_2(NCS)_2$.[12,13]

We present in Fig. 3 the film thickness dependence of the optical transmission spectra measured at 300 K. We can see that the thin films exhibit a strong and wide absorbance peak between 400 and 600 nm, which is in good agreement with previous results obtained on powder samples.[14] This main absorption peak exhibits features around 525, 485 and 445 nm that are clearly observable for thicknesses in the range 11-530nm, and result in the asymmetry of the band (due to a reduced signal-to-background ratio) for the 7 nm film. We therefore emphasize here that this absorption spectrum remains nearly unchanged even for the thinnest film (7 nm). This shows that the electronic structure of the films is almost the same as the thickness decreases.

These results demonstrate that it is possible to synthesize $Fe(phen)_2(NCS)_2$ thin films by high vacuum evaporation that are flat and maintain the expected electronic structure down to a thickness of ~10nm. In order to ascertain the spin transition, we have used a SQUID magnetometer to measure the temperature dependence of the magnetic susceptibility $\chi_m T$ of a powder sample and of a thin film (280 nm-thick on a silicon substrate) that is thick enough given this technique's sensitivity. [15] Referring to Fig. 4, we observe with increasing temperature a sudden rise of $\chi_m T$ at 175 K on both the powder and the thin film. As seen in the inset to Fig. 4, which shows the corresponding temperature dependence of the effective magnetic moment per molecule, this reflects the transition from a LS state (S=0; $\mu_{eff}$ ~0.5 $\mu_B$) to a HS state (S=2; $\mu_{eff}$ ~5 $\mu_B$). The spin transition of the thin film is more gradual than that of the powder sample, probably because of defects.[16]

We present in Fig. 5 the current-voltage (I-V) characteristics at room temperature of a device based on a 240 nm-thick $Fe(phen)_2(NCS)_2$ film deposited between gold electrodes. At low voltage, the ln(I)-ln(V) characteristic is linear with a slope of 1.17, which suggests that the conduction is ohmic.[17] Above 1.4 V, the ln(I)-ln(V) characteristic remains linear but with a slope of 2.04, indicating a space charge-limited current (SCLC) regime. In this regime, the current density is given by:[17]

$$j = \frac{9}{8}\mu\varepsilon\varepsilon_0 \frac{V^2}{L^3} \quad (1)$$

where $\mu$, $\varepsilon$, $\varepsilon_0$, V and L are respectively the carrier mobility, the dielectric constant, the vacuum permittivity, the applied voltage and the thickness of the film. The large (5.1 eV) work function of the gold electrodes generally favors hole injection into organic layers due to a low hole barrier height.[18] There are no reports on the HOMO position for this kind of SCO complexes. We can nevertheless deduce from the ohmic characteristic at low voltage that the HOMO energy level of $Fe(phen)_2(NCS)_2$ is likely near the work function of Au. From the SCLC relation (see Eq. 1), we thus deduce a hole mobility of $6.53\times10^{-6}$ $cm^2/V\cdot s$, which is similar to the lower range of mobility measured on the widely studied tris(8-hydroxyquinoline)aluminium.[19]

In conclusion, thin films of $Fe(phen)_2(NCS)_2$ with high uniformity and very low surface roughness have been grown using a high vacuum deposition technique. The films maintain the expected electronic structure down to at least 10 nm, and exhibit a spin transition at 175 K as evidenced by magnetic susceptibility measurements on a 280 nm-thick film. We extract a mobility of $6.53\times10^{-6}$ $cm^2/V\cdot s$ in the space charge-limited regime that is comparable to that found in typical small organic molecules. We aim in the future to perform temperature-dependent transport measurements, and moreover to develop a more sensitive method to probe the spin transition in ultrathin films. For instance, we expect that a $Fe(phen)_2(NCS)_2$ tunnel barrier can exist within a finite temperature range in two spin states with a different band gap, spin and charge scattering rates etc… This work paves the way for bistable multifunctional molecular devices based on spin-crossover complexes.

The authors thank A. Derory for SQUID measurements, and B. Doudin for access to the transport bench. This research was supported by the EC Sixth Framework Program (contract n°NMP3-CT-2006-033370) and the by the French national research agency (contract n°ANR-06-NANO-033-01).

**Figures caption**

Fig. 1 (color online) AFM images of $Fe(phen)_2(NCS)_2$ thin films deposited by high vacuum evaporation: (a) 280 nm on silicon substrate (2×2 $\mu m^2$), and (b) 240 nm on glass substrate (10×10 $\mu m^2$).

Fig. 2 XPS spectra for a 280 nm-thick $Fe(phen)_2(NCS)_2$ film at (a) the N 1*s* and (b) the Fe 2*p* edges. Inset of panel (a): the molecular structure of $Fe(phen)_2(NCS)_2$. Panel (b): the arrow indicates a shoulder that is characteristic of the HS state.

Fig. 3 (Color online) Optical transmission measured at room temperature for $Fe(phen)_2(NCS)_2$ thin films with varying thickness from 7 nm (top) to 530 nm (bottom).

Fig. 4 Temperature dependence of $\chi_m T$ for the powder and the 280 nm-thick film on silicon substrate. Inset: the temperature dependence of the magnetic moment per molecule.

Fig. 5 (Color online) Current-Voltage characteristics for a 240 nm-thick film of $Fe(phen)_2(NCS)_2$ at 300 K and linear fits to the data. Inset: device structure. The junction area is 2 μm×100 μm

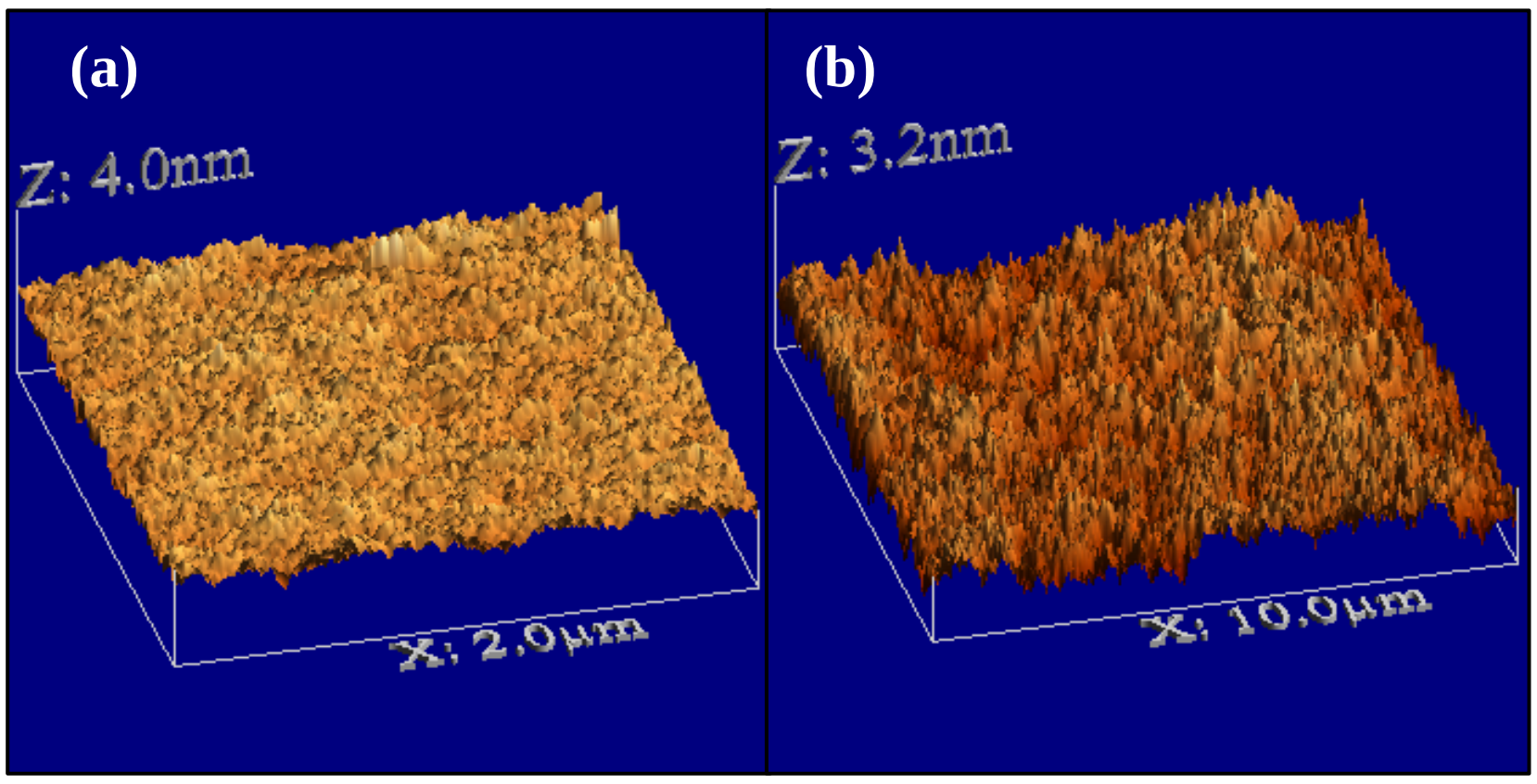

**Fig. 1 Shi *et al***

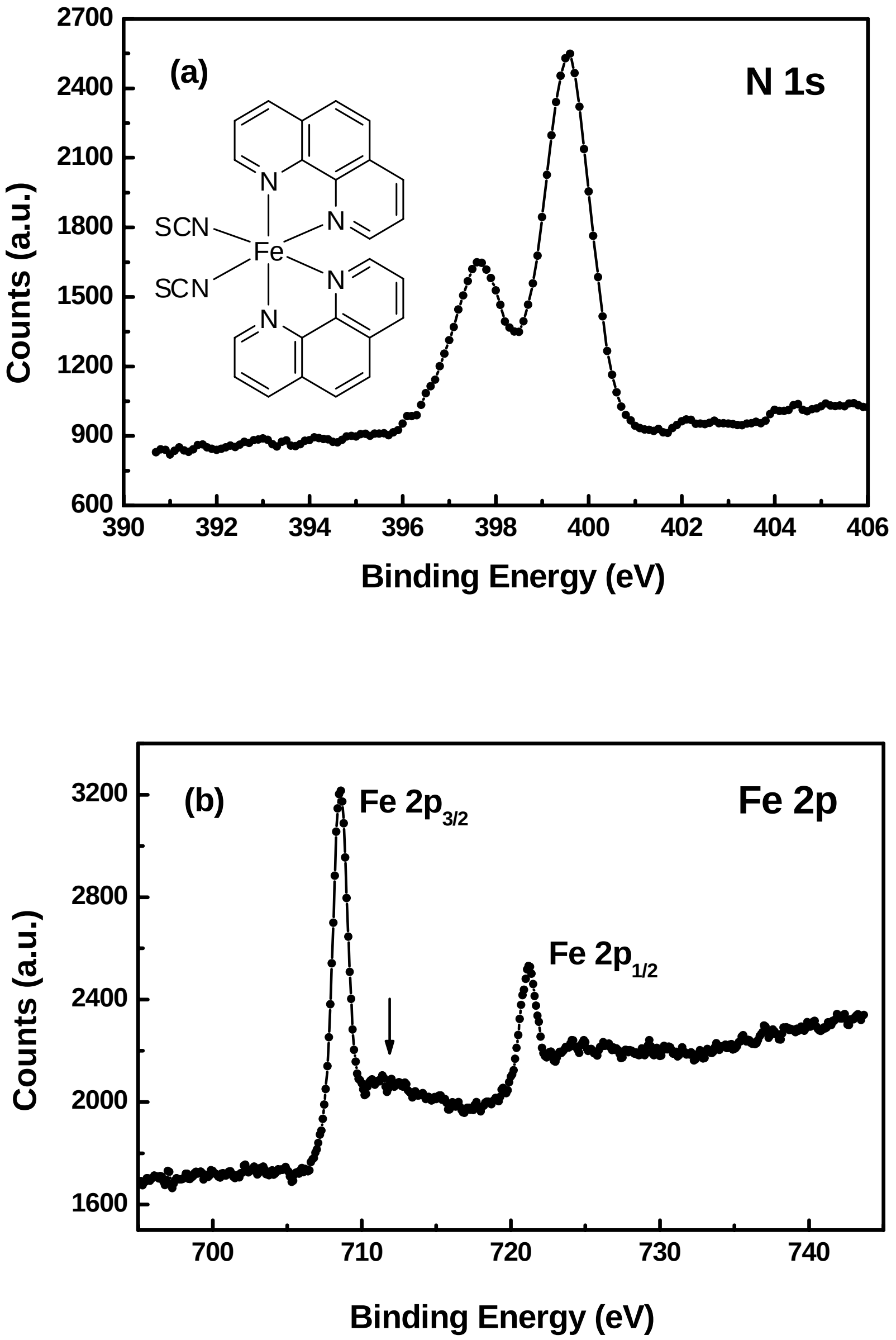

**Fig. 2** **Shi** ***et al***

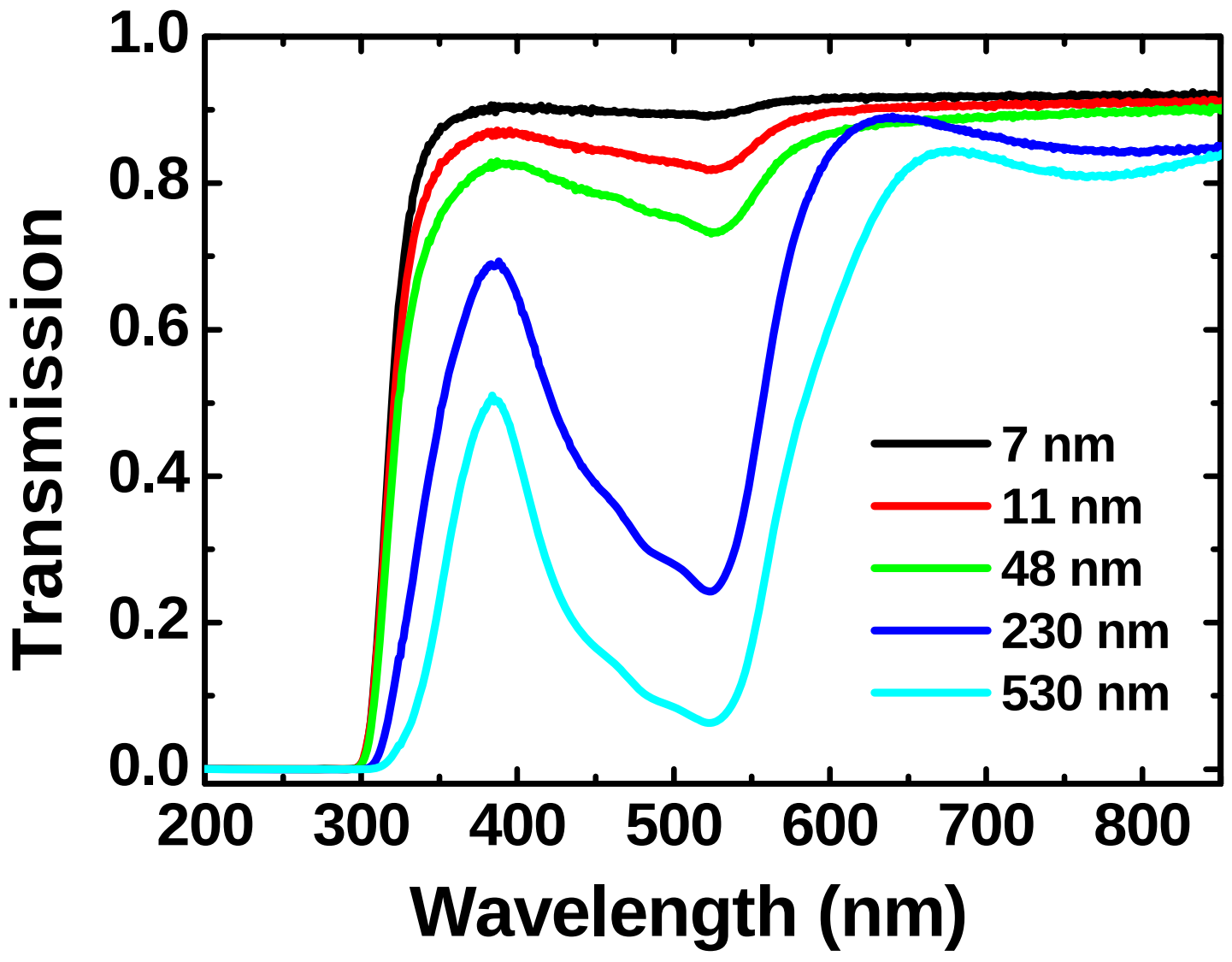

**Fig. 3 Shi *et al***

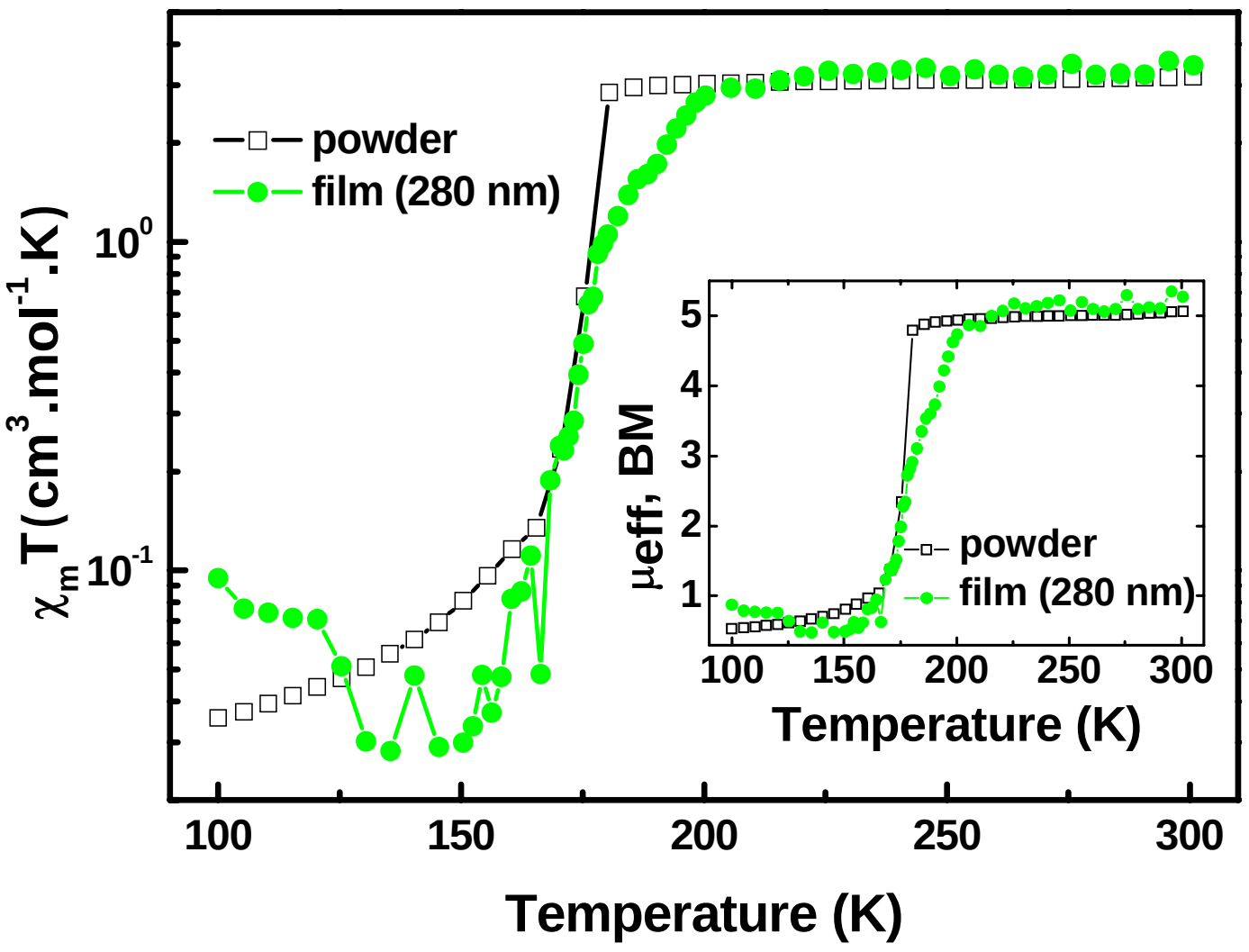

Fig. 4 Shi *et al*

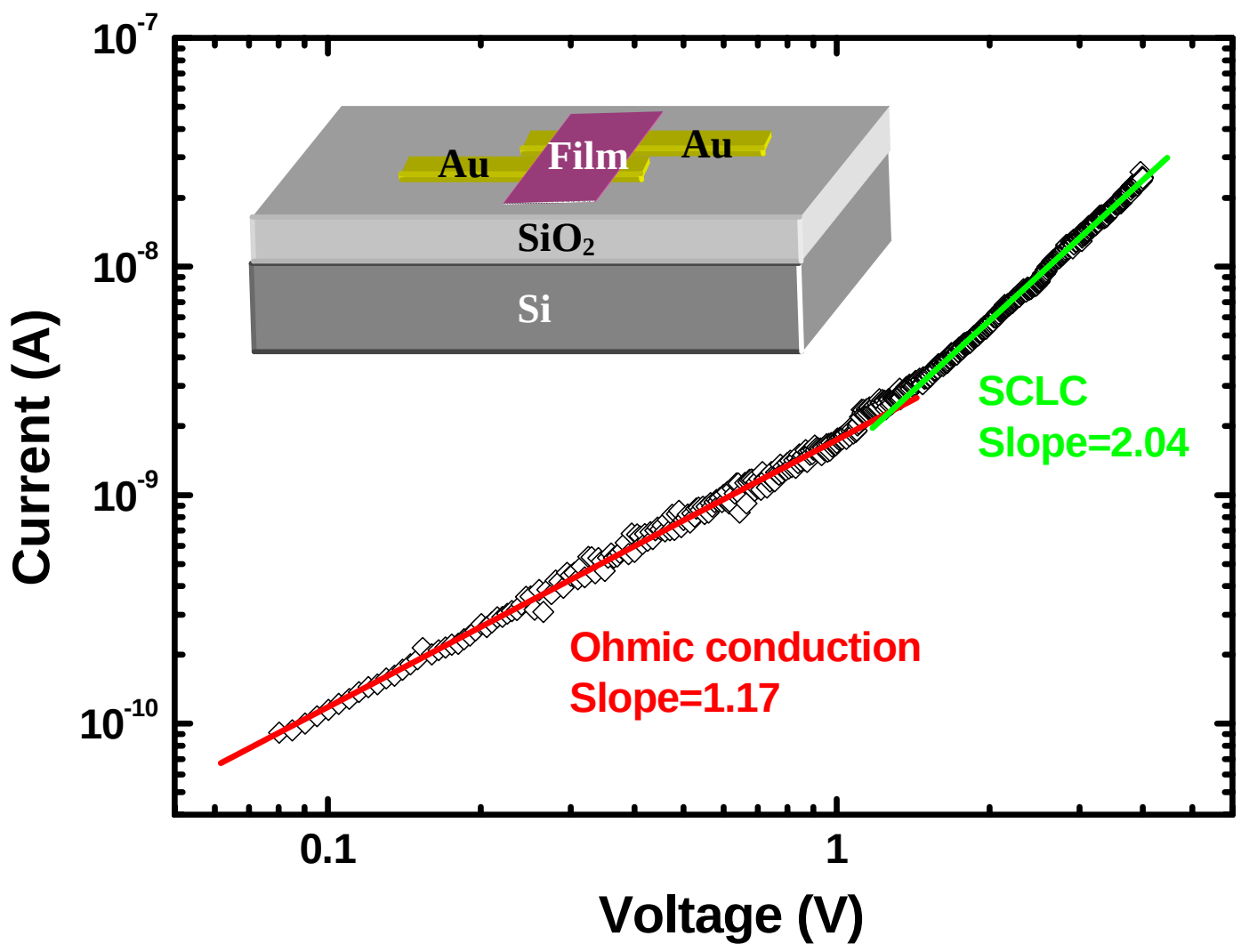

**Fig. 5 Shi *et al***